\documentclass[twocolumn,showpacs,preprintnumbers,amsmath,amssymb]{revtex4}
\usepackage{graphicx}
\usepackage{dcolumn}
\usepackage{bm}

\newcommand{\bef}{\begin{figure}}
\newcommand{\eef}{\end{figure}}
\newcommand{\nn}{\nonumber}
\newcommand{\be}{\begin{equation}}
\newcommand{\ee}{\end{equation}}
\newcommand{\bea}{\begin{eqnarray}}
\newcommand{\eea}{\end{eqnarray}}
\begin{document}

\title{Photon interferometry and size of the hot zone in relativistic heavy ion
 collisions}

\author{Jan-e Alam$^1$, Bedangadas Mohanty$^1$, Pradip Roy$^2$, Sourav 
Sarkar$^1$ and Bikash Sinha$^{1,2}$}

\medskip

\affiliation{$^1$Variable Energy Cyclotron Centre, Kolkata 700064, India\\
$^2$Saha Institute of Nuclear Physics, Kolkata 700064, India}

\date{\today}

\begin{abstract}
The parameters obtained from the theoretical analysis of the
single photon spectra observed by the WA98
collaboration at SPS energies have been used to evaluate the
two photon correlation functions. The single photon spectra and the two
photon correlations at RHIC energies have also been evaluated,
taking into account the effects
of the possible spectral change of hadrons in a thermal bath.  
We find that the ratio $R_{side}/R_{out}\,
\sim 1$ for SPS and $R_{side}/R_{out}\,<1$ for RHIC energy.

\end{abstract}

\pacs{25.75.+r,25.75.-q,12.38.Mh}
\maketitle

\section{INTRODUCTION}

  Two-particle intensity interferometry along with the analysis of 
  single-particle spectra has been widely used for a 
  quantitative characterization of the hot zone created by the 
  collision of two heavy ions at high energies. The method was first 
  applied to measure the angular diameter of stars and other 
  astronomical objects using the measurements based on two-photon 
 correlations known as Hanbury Brown-Twiss (HBT) 
correlations ~\cite{hb1}. Later in the field of particle physics, 
  two-pion correlations were used to obtain the spatial extent of 
  the fireball in proton - anti-proton reactions~\cite{hb2}. In high 
  energy heavy ion collisions two-particle ({\it e.g.} pions, kaons etc) 
  correlations have been 
  extensively studied both experimentally~\cite{na35,hb7,hb8}
 and theoretically~\cite{hb3,uaw}
 to obtain direct information about the size, 
  shape and dynamics of the source at freeze-out. This is 
  usually done via selection of transverse momentum and rapidity of 
  the correlated particles. Compared to two-particle correlation, 
  the three-particle correlation has been shown to provide additional 
  information on chaoticity and asymmetry of the source emission~\cite{hb4}. 
  Recently, it has been argued that spatial separation between sources
  of different species can be extracted from non-identical particle 
  correlation functions, hence these can provide an independent cross-check
  of the transverse flow prescription~\cite{hb5}. One of the major
  limitations of carrying out the correlation studies 
  with hadrons appearing at the final state is that, the information 
  about the possible early dense state of matter is diluted or lost through
  re-scattering. 
  Although, some calculations suggest that the ratio of the radius of the 
  source along the direction of the total transverse momentum ($\vec{K}_T$) 
  of the two detected particles to the source radius perpendicular to 
  $\vec{K}_T$  and the beam direction,
  as a function of transverse momentum can provide useful 
  information~\cite{hb6}. However, the results at relativistic 
  heavy ion collider (RHIC) where quark-gluon plasma (QGP) is expected
  to be formed does not show any encouraging result on the above 
  signal~\cite{hb7,hb8}. Investigations
  to understand the partonic effects on the interferometry with 
  final state particles (pions) at RHIC~\cite{hb9} are being pursued. 
  HBT interferometry as a sensitive probe
  of the QCD equation of state and hence formation of QGP
  has been discussed in~\cite{hb6,hb9,hb10}.
  It has been argued~\cite{dks,timmermann}  that in contrast 
  to hadrons, two-particle intensity 
  interferometry of photons which are produced throughout the space-time 
  evolution of the reaction and which suffer almost no interactions with 
  the surrounding medium can provide information on the 
  the history of the evolution of the hot 
  matter created in heavy ion collisions. 

  From the experimental point of view, photon interferometry encounters
  considerable difficulties compared to hadron interferometry due 
  to small yield of direct photons from the early hot and 
  dense region of the matter and the associated large background of photons 
  primarily from the electro-magnetic decay processes of hadrons at freeze-out
  ~\cite{wa98}. However, recent calculations demonstrate that it is still 
  possible to experimentally 
  filter out the correlations of photons from the hot and dense zone from 
  those arising due to the residual correlations of decay photons. This can
  be done by studying photon 
  interferometry as a function of invariant relative momentum of the two 
  photons~\cite{peressounko}. One expects the contribution from the 
  photons coming from the early stage of the reactions to contribute 
  predominantly to the region of small invariant relative momentum. 

  The aim of the  present work is to first analyze the
single photon spectra observed  by the WA98 
collaboration~\cite{wa98} in Pb + Pb interactions at SPS energies.
In the next step we evaluate the two photon correlations with 
the initial conditions, equation of states (EOS) and freeze-out
conditions which reproduce the  photon spectra mentioned
above. The EOS used here~\cite{bm2} contains all hadrons upto
mass $\sim 2.5$ GeV~\cite{pdg}. The freeze-out conditions  are fixed from the
study of hadronic spectra measured by NA49 collaboration~\cite{na49}.
The effects of the possible spectral modifications of hadrons in a thermal
bath on the single photon spectra and two photons correlations 
have been included. The results at RHIC energy for the single photon 
distributions and two photon correlations are also presented.  

It is necessary to point out here that our calculation 
contains some significant improvements over the previous 
works~\cite{dks,timmermann}.
The correlation function evaluated here (i) contain  contributions 
from two loop calculations of photon emission rate~\cite{pa}, (ii)
emission rate from hadronic matter contain in-medium effects, (iii)
the EOS contains all hadronic degrees of freedom upto mass
$\sim 2.5$ GeV and (iv) the initial conditions, freeze-out 
conditions and the EOS reproduces the WA98 single photon
spectra~\cite{wa98} and NA49 hadronic spectra.
Moreover, in Ref.~\cite{timmermann} where the 
correct definition of the BEC was used, the space time evolution
was only one-dimensional.  

 The paper is organized as follows. In the next section we give a
 general discussion on the correlation function and 
 the associated kinematics. This is followed by the section which 
 deals with the space time evolution. 
 Section~IV, deals with the results. First 
 we present the results for SPS energies. Then we discuss the 
 single photon spectra and two photon correlations 
 at RHIC energies. In section~V we summarize our findings.

\section{CORRELATION FUNCTION}

The Bose-Einstein correlation (BEC) function for two identical particles
is defined as,
\begin{equation}
C_{2}(\vec{k_{1}}, \vec{k_{2}}) = \frac{P_{2}(\vec{k_{1}}, \vec{k_{2}})}
{P_{1}(\vec{k_{1}}) P_{1}(\vec{k_{2}})} 
\label{eq1}
\end{equation}
where $\vec{k_i}$ is the three momentum of the particle $i$ and 
$P_{1}(\vec{k_{i}})$
and $P_{2}(\vec{k_{1}}, \vec{k_{2}})$ represent 
the one- and two- particle inclusive
photon spectra respectively. These are defined as,

\begin{equation}
P_{1}(\vec{k}) = \int d^{4}x~\omega (x,k)
\label{eq2}
\end{equation}
and
\begin{widetext}
\begin{eqnarray}
P_{2}(\vec{k_{1}}, \vec{k_{2}}) = P_{1}(\vec{k_{1}}) P_{1}(\vec{k_{2}})  +
\int d^{4}x_{1} d^{4}x_{2} ~\omega (x_{1},K) 
~\omega (x_{2},K)~\cos(\Delta x^{\mu} \Delta k_{\mu}) 
\label{eq3}
\end{eqnarray}
\end{widetext}
where, $K=(k_1+k_2)/2$, $\Delta k_\mu=k_{1\mu}-k_{2\mu}=q_\mu$,
$x$ and $k$ are the four-coordinates of position and momentum 
respectively and $\omega(x,k)$ is the source function, which  defines
the average number of particles with four-momentum $k$ emitted from a 
source element centered at the space-time point $x$. $\omega(x,k)$
is actually the thermal emission rate of photons per unit four volume. 
For the QGP phase at a temperature $T$ this is given by~\cite{jk,rb,pa},

\begin{widetext}
\begin{equation}
\omega(x,k) = \frac{5}{9}\,\frac{ \alpha \alpha_{s}}{2 \pi^{2}} 
T^{2}(x) e^{-k/T(x)}
\Bigl[ \ln \Bigl[ \frac{2.912 k}{g^{2} T(x)} \Bigr] + \frac{4(J_{T} - J_{L})}{\pi^3} \Bigl(\ln 2 + \frac{k}{3T(x)} \Bigr) \Bigr] 
\label{eq4}
\end{equation}
\end{widetext}
where, $\alpha$ = 1/137, $\alpha_{s}$ is the
strong coupling constant, 
$J_{T}$ = 4.45 and $J_{L}$ = -4.26. Note that 
$C_2$ is independent of the strong coupling constant.
For photon emission from a hot hadronic gas, we have considered a host
of reaction processes involving $\pi$, $\rho$, $\eta$ and $\omega$ 
mesons as well as from the decay of the $\rho$ and $\omega$~\cite{jk}. 
For the details on the evaluations of the photon emission rate
form a hadronic gas we refer to~\cite{npa1,npa2}. The effects
of the intermediary $a_1$ meson have also been considered in evaluating
the photon emission rate~\cite{kim,annals}.

We shall be presenting the results in terms of difference in 
rapidity ($\Delta y$) of the two photons,
outward ($q_{out}$), side-ward ($q_{side}$) 
and invariant momentum differences($q_{inv}$) of the two photons 
and these are defined as,

\bea
q_{out}& = & \frac{\vec{q}_{T} \cdot \vec{K}_{T}}{\left|{K_{T}}\right|}\nn\\
       & = & \frac{(k_{1T}^{2} - k_{2T}^2)}{
\sqrt{k_{1T}^{2} + k_{2T}^{2} + 2 k_{1T}k_{2T} \cos(\psi_{1}-\psi_{2})}}
\label{eq5}
\eea
\bea
q_{side}& = &{\left| \vec{q}_{T} - q_{out} \frac{ \vec{K}_{T}}{K_{T}}\right|}
\nn\\ 
	& = &\frac{2 k_{1T}k_{2T} \sqrt{1-\cos^{2}(\psi_{1}-\psi_{2})}}{
\sqrt{k_{1T}^{2} + k_{2T}^{2} + 2 k_{1T}k_{2T} \cos(\psi_{1}-\psi_{2})}}
\label{eq6}
\eea
\bea
q_{inv} & = &\sqrt{-2 k_{1T}k_{2T} \Bigl[\cosh(y_{1}-y_{2}) - \cos(\psi_{1}
-\psi_{2}) \Bigr]},\nn\\
\label{eq7}
\eea
where, $\vec{q}_{T} = \vec{k}_{1T} - \vec{k}_{2T}$, $\vec{K}_{T} =
(\vec{k}_{1T} + \vec{k}_{2T})/2$ with the subscript $T$ indicating
the transverse component,
$y_i$ is the rapidity and $\psi_i$'s are the angles made 
by $k_{iT}$ with the $x$-axis. One also defines another 
kinematic variable $q_{long}$ as
\bea
q_{long} & = &k_{1z}-k_{2z}\nn\\
	 & = &k_{1T}\sinh(y_1)-k_{2T}\sinh(y_2)
\label{eq8}
\eea

It may be mentioned that the BEC function has values $1~\le~C_{2}(\vec{k}_{1},
\vec{k}_{2})~\le~2$ for a chaotic source. These bounds are from quantum 
statistics. While the radius corresponding to $q_{side}$ ($R_{side}$)
is closely related to
the transverse size of the system, the radius corresponding to $q_{out}$ 
($R_{out}$) measures both the transverse size and duration of particle 
emission~\cite{hb3,uaw}. 
So studying $R_{side}$/$R_{out}$ will indicate about the 
duration of particle emission~\cite{hermm,chappm,hb11}.
These source dimensions can be obtained by parametrizing the calculated 
correlation function with the empirical Gaussian form
\be
C_2=1+\exp(-R^2_{out}q^2_{out}-R^2_{side}q^2_{side}-R^2_{long}q^2_{long})
\label{param1}
\ee
A gross idea of the source size can also be inferred from $R_{inv}$ which
is defined as
\be
C_2=1+\exp(-R^2_{inv}q^2_{inv})
\label{param2}
\ee 


\section{Space-time evolution}
To compare transverse momentum spectra of photon
with  the experiments we have to convolute the
static emission rate (fixed temperature, e.g. given by
eq.~\ref{eq4}) by the space-time evolution from the
formation to the freeze-out state.
This is done solving the  relativistic hydrodynamical
equations in (3+1) dimension~\cite{hvg} with boost invariance along
the longitudinal direction~\cite{jdb}. 
One then needs to specify the following: the
initial temperature and time when the hot system reaches a state of 
thermodynamic equilibrium, the EOS which
guides the rate of expansion/cooling, and the freeze-out temperature
when the system decouples into free-streaming hadrons.
We will not discuss here  the details on the EOS, initial conditions 
and the modification of the hadronic masses in a thermal bath because
these topics had been discussed in our earlier works in great detail,
which will be appropriately referred below.
For SPS conditions we will assume a hot hadronic gas in the initial
state with temperature 
dependent masses of the constituent hadrons which expands and
cools till freeze-out. The two-photon correlations at SPS energies
will also be  given for a possible phase transition scenario
{\it i.e.} when QGP is formed at the initial state.
The EOS used 
here is given in ~\cite{bm2}. The initial conditions are similar
to ref.~\cite{prcr,reds}. The variation of hadronic masses (except 
pseudoscalar) with temperature follows universal scaling law
proposed by Brown and Rho (BR)~\cite{geb}.
It is necessary to point out that a substantial amount of literature 
is devoted to the calculation of effective masses of hadrons
in the medium. Our principal motivation for choosing the BR
conjecture is that this has been used successfully to explain
the WA98 photon spectra~\cite{prcr} and CERES/NA45~\cite{ceres} 
low mass dilepton spectra~\cite{liko,ss}.
For higher collision energies, i.e. at RHIC we will 
consider a situation where QGP is formed in the initial state, 
then the quark matter
evolves with time to a hadronic phase via an intermediate mixed phase in a
first order phase transition scenario. The mixed phase is a mixture of both 
quark matter and hadronic matter, with the fraction of quark matter decreasing
with time to zero when the phase transition is complete.
The hot hadronic gas then expands till the system freezes-out.
The initial condition for the SPS and 
RHIC energies in terms of the initial temperature ($T_{i}$) is set from 
the number of particles per unit rapidity at the mid-rapidity region for those 
energies according to the following equation,
\be
T_i^3=\frac{2\pi^4}{45\zeta(3)}\frac{1}{\pi R_A^2\tau_i 4a_k}\,
\frac{dN}{dy}
\ee
$a_k=\pi^2g_k/90$ is determined by the statistical degeneracy ($g_k$)
of the system formed after the collision.
Taking the particle multiplicity per unit rapidity 
at mid rapidity to be 700 for SPS, we get $T_{i}$ = 200 MeV with
the formation time $(\tau_i)$ taken as 1.0 fm/c.  
With a multiplicity of 1100 for Au+Au collisions at RHIC, 
we get $T_{i}$ = 264 MeV for an initial time  
of 0.6 fm/c~\cite{dumitru}. The critical temperature ($T_{c}$) 
is taken to be 170 MeV~\cite{karsch} here.
The freeze-out temperature ($T_{f}$) 
is taken to be 120 MeV for both SPS and RHIC energies; 
a value that describes the transverse momentum 
distributions of hadrons produced~\cite{bkp}. 
Let us now turn to the EOS which plays a central role in
the space time evolution, we have considered.
For the hadronic phase the EOS corresponds to that for the 
hadronic gas with particles of mass up to
2.5 GeV and includes the effects of non-zero widths of various mesonic 
and hadronic degrees of freedom~\cite{bm2}. 
The velocity of sound ($c_{s}^{2}$) corresponding to this EOS at freeze-out 
is about 0.18~\cite{bm2} (the value corresponding to a free gas of massless 
particles being 0.33). We emphasize that the medium modifications of the 
constituent hadrons play a non trivial role in the estimation of 
the quantities mentioned above. Of special mention is the velocity of
sound and the estimation of the initial temperature for the hot hadronic
gas considered at SPS. These have been exhaustively dealt with in a
number of our earlier works~\cite{bm2,annals,prcr,bkp} 
and we do not repeat them here. 
With these inputs we have performed a (3+1)-dimensional hydrodynamic expansion.
The solution of the hydrodynamic equations has been used to evaluate 
the space time integration involved in Eqs. ~\ref{eq2} and \ref{eq3}.

\section{Results}

In this section we present the results of two-photon interferometry. We 
shall first present the results for SPS energies and then give our predictions
for RHIC. In both the cases, we will first start with the single photon 
spectra and then present the two-photon correlation as a function of 
$\Delta y$, $q_{out}$, $q_{side}$ and $q_{inv}$.
At the SPS energies we present the two-photon correlation results 
using the inputs which reproduces the  the measured 
single photon spectra~\cite{wa98} as mentioned before. 
For RHIC our predictions
are based on the same model as that explains the SPS data, 
but for different initial conditions as expected at RHIC energies. 

\bef
\begin{center}
\includegraphics[scale=0.4]{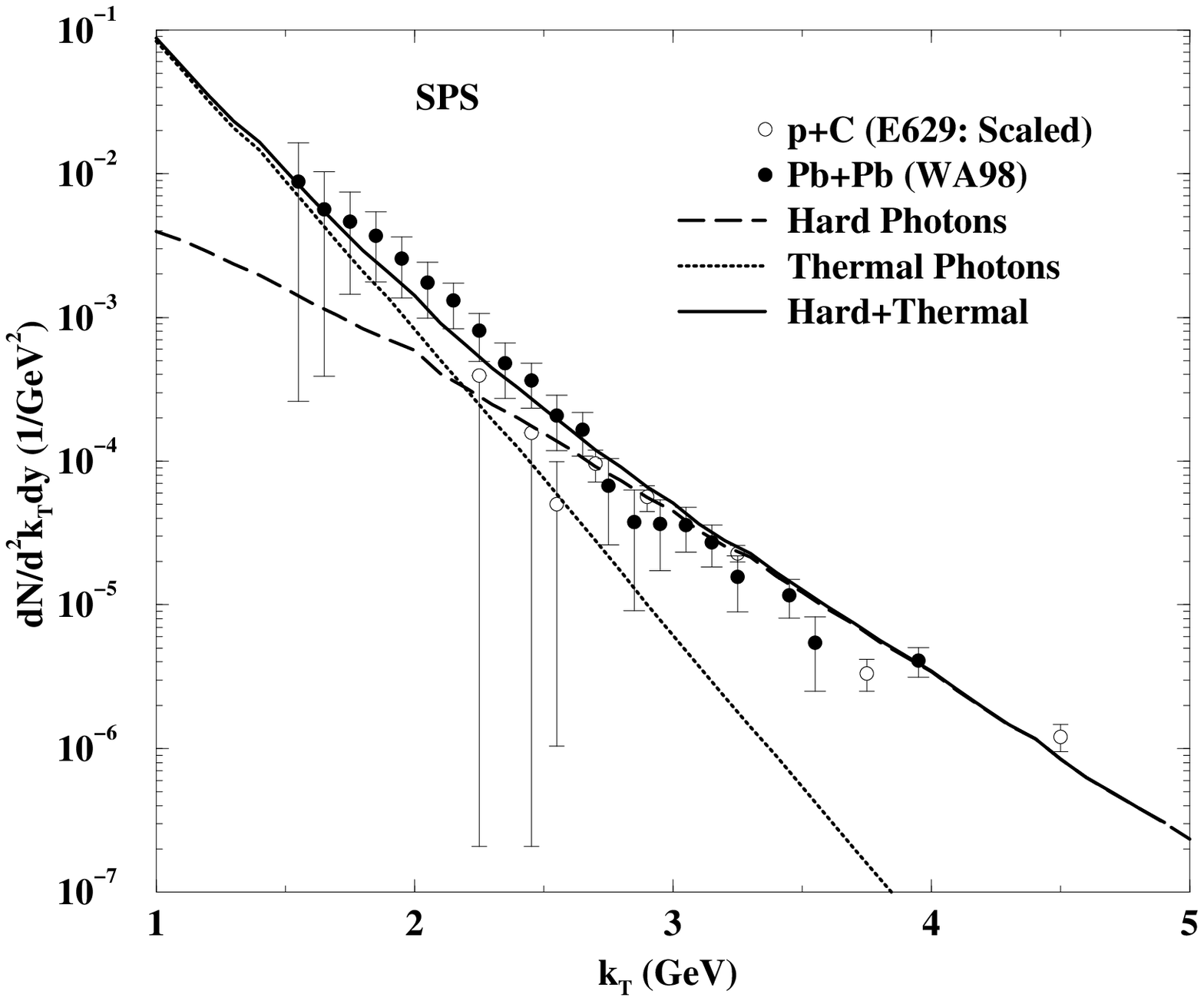}
\caption{Single photon spectra at SPS
}
\label{fig1}
\end{center}
\eef

Fig.~\ref{fig1}, shows the single photon yield measured by the WA98 
experiment at CERN SPS, as a function of transverse momentum. 
The contributions from hard scattering of partons (hard photons) and
from a hot hadronic gas (thermal photons)
have been shown separately. The hard photon contribution
has been normalized to reproduce the scaled
p+C data~\cite{e629} as well as the scaled p-p data~\cite{e704}
(p-p data is not shown here. Please see ref.~\cite{prcr}).
The experimental results of WA98 collaboration show some
``excess'' photons over the hard photons for 
$1.5\le\,p_T$ (GeV)\,$<2.5$ GeV, which
we argue to originate from a thermal source either of QGP
or hot hadronic gas of initial temperature 200 MeV
(see also ~\cite{prcr,reds} for details). In the present
work we have evaluated the thermal photon from a hot hadronic
gas with their mass reduced according to the universal scaling 
scenario~\cite{geb}.
The combined results of hard and thermal photons
is found to explain the data reasonably well. 

From the expressions given in Eqs.~\ref{eq5}\,,\ref{eq6}\,,\ref{eq7} 
one can see that several 
combinations can be made in the variables $y$, $\psi$ and $k_{T}$ to present
the results for two-photon correlation.  For simplicity, we will present all 
two-photon correlation results for direct photons with momentum around 2 GeV/c
since it lies in the range of momentum where excess direct photons
have been observed at SPS energies. We take $\psi_{2} = 0$ and $y_{2} = 0$
for all cases. We vary $\psi_{1}$ and $y_{1}$ wherever necessary.

\bef
\begin{center}
\includegraphics[scale=0.5]{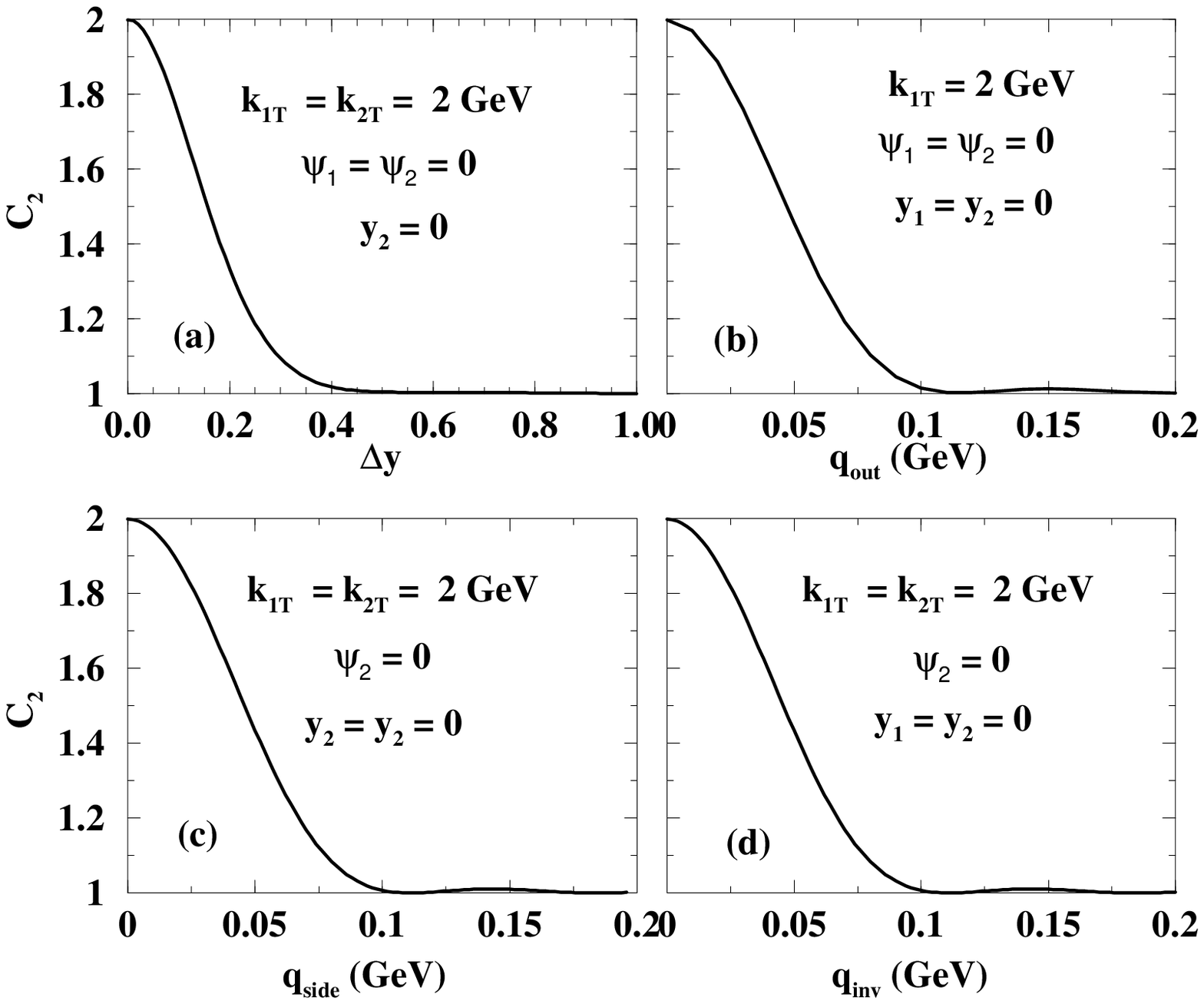}
\caption{Correlation function $C_2$ as a function of $\Delta y$,
$q_{out}$, $q_{side}$ and $q_{inv}$
for Pb+Pb collisions at 158 AGeV at SPS}
\label{fig2}
\end{center}
\eef
Figs.~\ref{fig2}(a-d) show the variation of the correlation 
strength ($C_{2}$) as a function of $\Delta y$, $q_{out}$, $q_{side}$
and $q_{inv}$ for various phases (sum $\equiv$ QGP+Mixed+hadrons). 
The HBT radii of the evolving
hot hadronic matter (denoted by Hadron$^*$ indicating that the 
masses of the hadrons are modified in the medium according to BR
scaling) is shown in Table~1. Please note that to get the quantity
$R_{long}$ in the longitudinally co-moving system (LCMS)of reference
one should multiply the numbers given in Table I by
$cosh(y_K)$ where $y_K$ is the rapidity corresponding to the
momentum $K$ defined in section II.
We would like to mention here that the HBT radii give the length
of homogeneity of the source and this is equal to the geometric 
size if the source is static. However, for a dynamic source 
{e.g.} the system formed after ultra-relativistic heavy ion 
collisions, the HBT radii is smaller than the geometric 
size (see ~\cite{chappm,csorgo,bertsch,xu}).
\bef
\begin{center}
\includegraphics[scale=0.5]{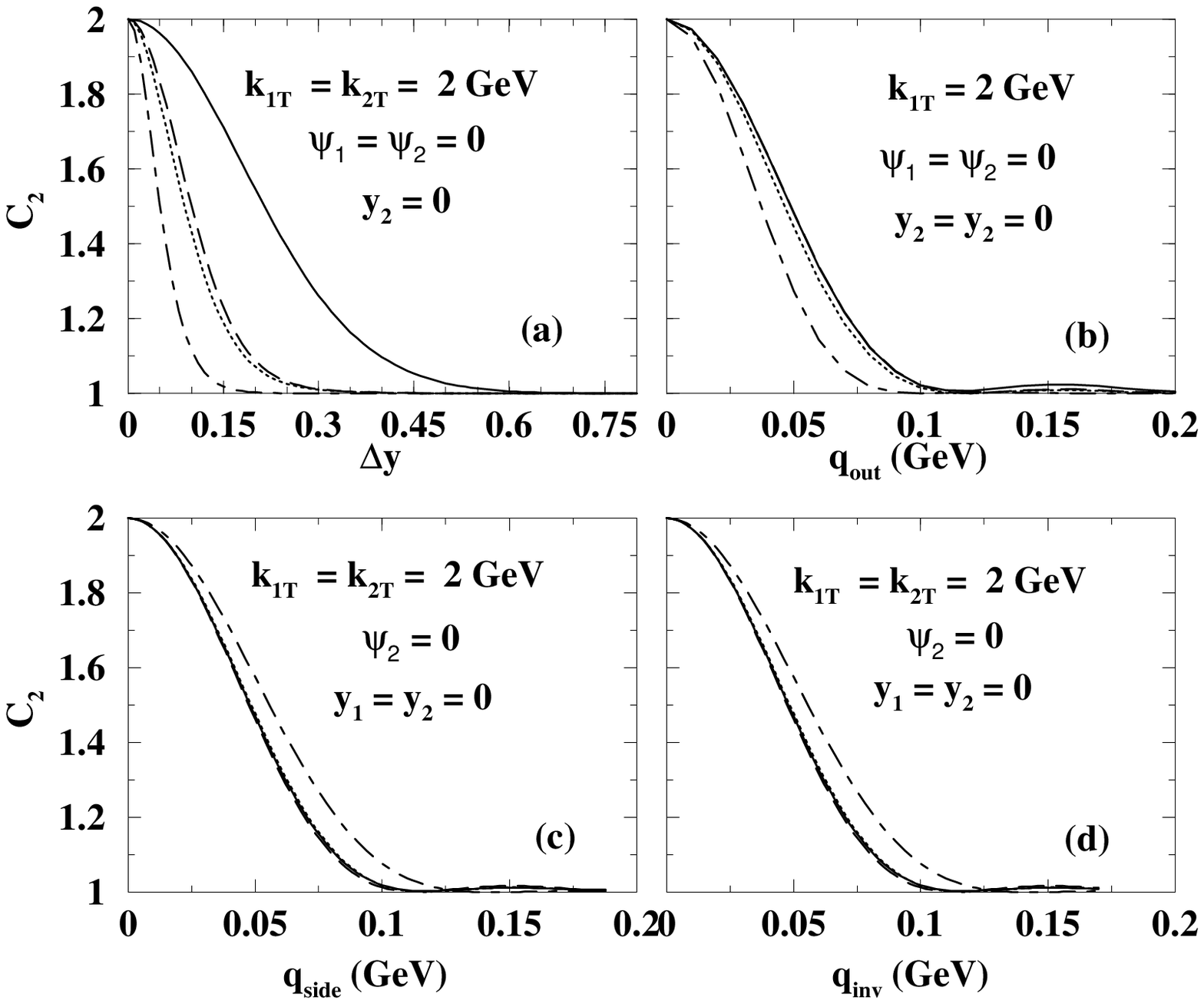}
\caption{Correlation function, $C_2$ as a function of $\Delta y$,
$q_{out}$, $q_{side}$ and $q_{inv}$
for Pb + Pb  collisions at 158 AGeV at SPS. Solid (dashed)
line indicate results for QGP (Mixed) phase and dot-dashed (dotted) line 
represent correlation function for hadronic (sum) phase.
}
\label{fig2b}
\end{center}
\eef

\bef
\begin{center}
\includegraphics[scale=0.4]{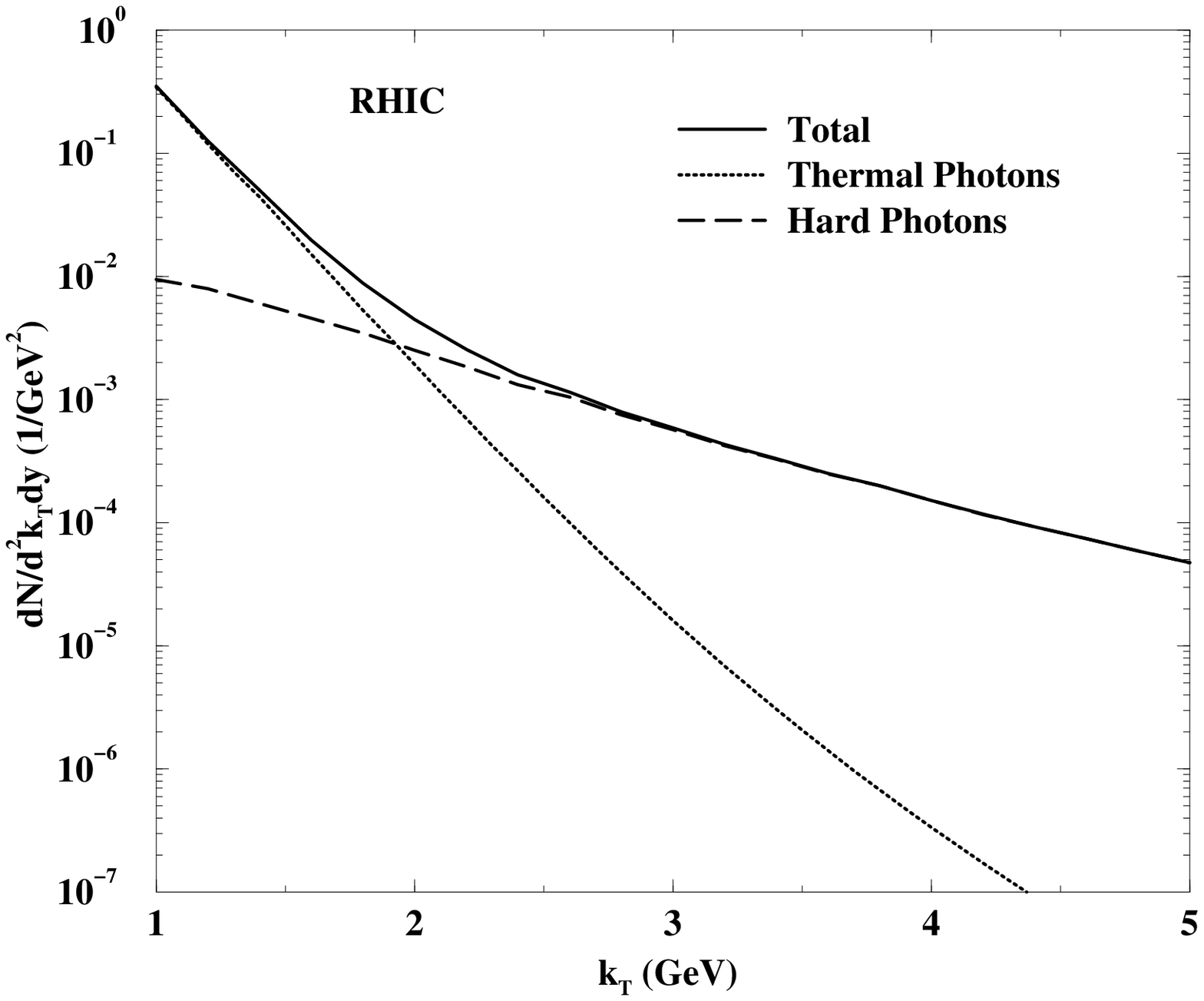}
\caption{Single photon spectra from Au+Au collisions at RHIC energies.
}
\label{fig3}
\end{center}
\eef
Our earlier analysis~\cite{prcr} shows that the WA98 single photon
spectra can be explained by two different kind of initial
states: (i) hot hadronic gas, with the masses of the hadrons 
vary according to BR scaling hypothesis 
or (ii) QGP is created initially.
It is rather difficult to distinguish between the two at present.
Therefore, in the following we show the results on 
two-photon correlation functions for case (ii) also.
In Fig.~\ref{fig2b} the variation of the two-photon correlations 
is depicted in a scenario when QGP is formed initially at SPS. The initial
conditions here are similar to the that of Ref.~\cite{prcr}.
The HBT dimensions extracted from these correlation functions
are shown in Table I.
The difference in the correlation function as a function 
of $q_{out}$ for the QGP and the mixed phase is negligible.    
When plotted as a function of $q_{side}$ (Fig.~\ref{fig2b}b)
the correlation function for the QGP, mixed phase and the sum
almost overlap. The width of the correlation function for the
hadronic phase is the largest as compared to the other
phases. The HBT dimensions satisfy the relation 
$R_{side}/R_{out}\,\sim\,1$ for
the correlation functions denoted by `sum' in the Table I.
This is irrespective of the formation of the  QGP or
hot hadronic gas with the mass of the hadrons (except pseudo-scalar)
approaching zero at the initial temperature. Although the later
scenario is closely related to the chiral/deconfinement
phase transition it is very difficult at this stage to draw a firm
conclusion.

Fig.~\ref{fig3}, shows our predictions for the single photon spectra at RHIC
energies. Here the thermal photons contain contribution from QGP,
mixed and the hadronic phases. The results show a clear dominance of
the thermal photon over the hard photons for $p_T<3$ GeV.	
We calculate
the two-photon correlation at RHIC energies with the inputs that
are used to evaluate the single photon spectra as a function of $\Delta y$,
$q_{out}$, $q_{side}$ and $q_{inv}$.

\bef
\begin{center}
\includegraphics[scale=0.5]{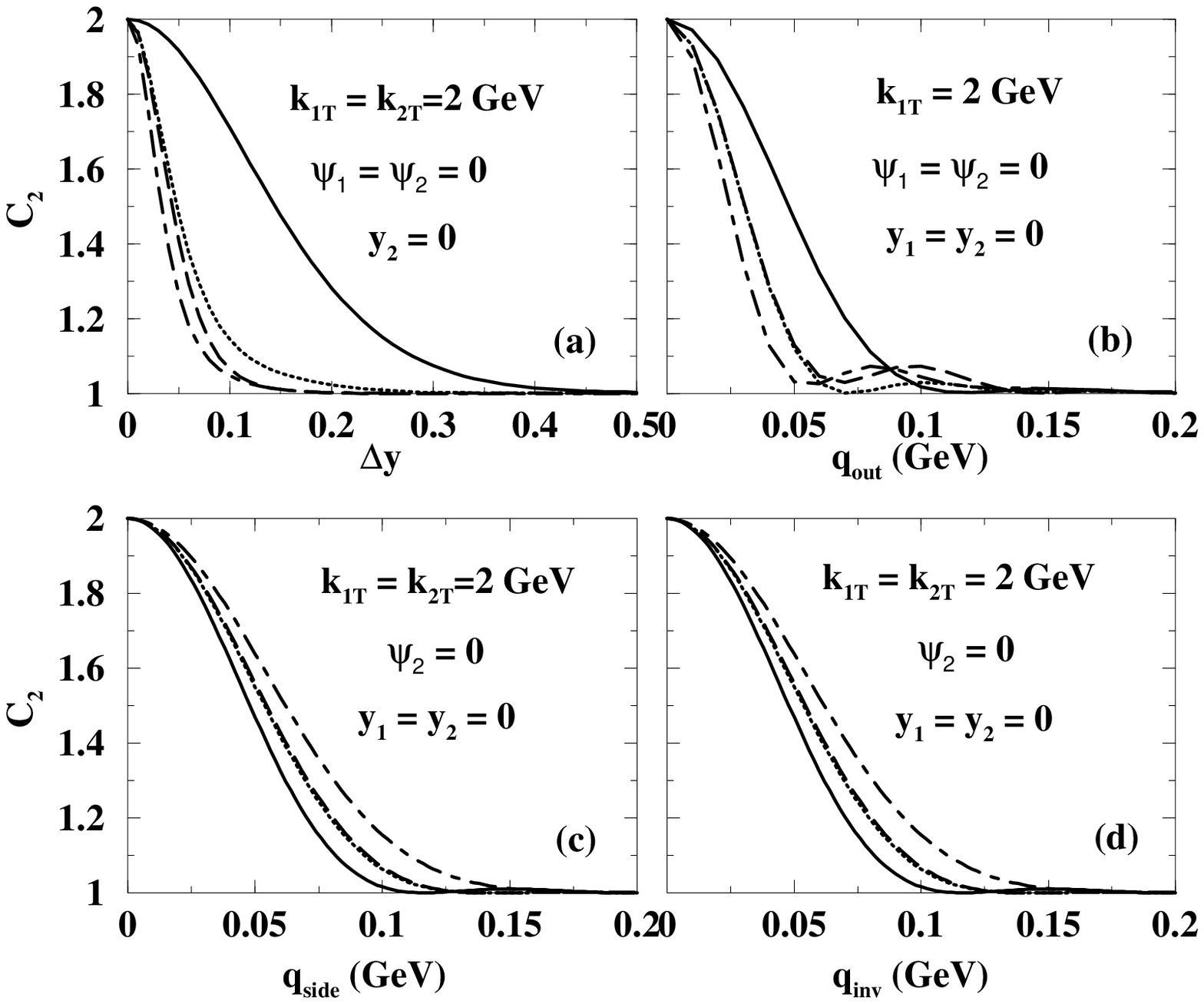}
\caption{Correlation function $C_2$ as a function of $\Delta y$,
$q_{out}$, $q_{side}$ and $q_{inv}$
for Au+Au collisions at 200 AGeV at RHIC. Solid (dashed) line indicate
results for QGP (Mixed) phase and dot-dashed (dotted) line 
represent correlation function for hadronic (sum) phase.
}
\label{fig4}
\end{center}
\eef
\renewcommand{\arraystretch}{1.5}
\vskip 0.2in
\begin{center}
\begin{tabular}{|c|c|c|c|c|c|c|}
\hline
& & $R_{inv}$ & $R_{out}$ & $R_{side}$ & $R_{long}$  \\
&  &   &  &     &   \\
\hline
SPS & Hadron$^*$ & 3.7 & 3.6 & 3.7 & 0.5\\
 &  &  &  &  & \\
\hline
& QGP &3.5& 3.4 & 3.5 &0.37\\
SPS & Mixed & 3.5 & 3.4 &3.5 & 0.8\\
& Hadron & 3.0 & 4.5 & 3.0 & 1.5\\
& Sum & 3.5 & 3.6 & 3.5 & 0.9\\
\hline
& QGP & 3.5 & 3.5 & 3.5 &0.55\\
RHIC & Mixed & 3 & 5.4 & 3 & 2.0\\
& Hadron & 2.7 & 6.8 & 2.7 & 2.3\\
& Sum & 3 & 5.5 & 3& 1.6\\
\hline
\end{tabular}
\end{center}
Table 1 : Values of the various parameters (in fm)
of the correlation functions in the forms given in Eqs.
\ref{param1} and \ref{param2} for SPS and RHIC energies at a transverse
momentum 2 GeV. 

Fig.~\ref{fig4}a shows the variation of $C_{2}$ as a function of $\Delta y$, for
two photons with momentum of 2 GeV/c. We have taken $\psi_{1}~=~\psi_{2}~=0$ 
for simplicity. The contribution from all the three phases are shown. 
We observe that the width of the correlation function is  
largest for the QGP phase followed by that from the mixed phase and then
the hadronic phase. The inverse will reflect the corresponding
HBT radii of the source.
Fig.~\ref{fig4}b shows the variation of $C_{2}$ with $q_{out}$, for 
$y_{1}~=~y_{2}~=~0$, $\psi_{1}~=\psi_{2}~=0$ and $k_{1T}$ = 2 GeV.
The contribution of all three phases are shown. One observes that the
$C_{2}$ variation is similar for the three phases and is similar to that for
the case with $\Delta y$ but the widths are considerably smaller.
Fig.~\ref{fig4}c shows how $C_{2}$ varies with $q_{side}$ for all the
three phases as well as the sum. For this case we have taken,
$y_{1}~=~y_{2}~=~0$, $\psi_{2}~=0$ and $k_{1T}~=~k_{2T}~=~2$ GeV.
The trend in the variation of the width for the three phase has reversed.
The width of $C_{2}$ distribution for QGP is smallest, followed by that
for the mixed phase and then the hadronic phase. 
Fig.~\ref{fig4}d shows how $C_{2}$ varies with $q_{inv}$ for all the
three phases as well as the sum. For this case we have taken,
$y_{1}~=~y_{2}~=~0$, $\psi_{2}~=0$ and $k_{1T}~=~k_{2T}~=~2$ GeV.
The trend in the variation of the width for the three phases is similar 
to that for the case of $q_{side}$.

\section{SUMMARY}
The two photon correlation functions has been evaluated for
SPS and RHIC energies.  Constraints from the experimentally 
observed single photon spectra has been used to evaluate the 
correlation functions at SPS energies. Predictions for both
the single photon spectra and the two photon correlation functions have 
also been given at RHIC energies. The values of HBT radii extracted 
from the two-photon correlation functions show that 
$R_{side}/R_{out}\,\sim 1$ irrespective of the formation 
of QGP or hadronic gas with reduced mass initially
for SPS and $R_{side}/R_{out}\,<1$ for RHIC energy.

{{\bf Acknowledgments}: (One of us (B.M.) is grateful to the Board of Research
on Nuclear Science and Department of Atomic Energy,
Government. of India for financial support in the form of Dr. K.S. Krishnan
fellowship.
}

\end{document}